\documentclass{article}
\usepackage{amsmath}
\usepackage{amsfonts}
\usepackage{amssymb}
\usepackage{graphicx}
\usepackage{setspace}
\usepackage[spanish]{babel}

\ifx\pdfoutput\@undefined\usepackage[usenames,dvips]{color}
\else\usepackage[usenames,dvipsnames]{color}
\IfFileExists{pdfcolmk.sty}{\usepackage{pdfcolmk}}{} 
\fi
\usepackage[plainpages=false,pdfpagelabels,pagebackref=false,naturalnames=true,hyperindex=true,pdftitle={Un metodo efectivo y estable para la evaluacion de la complejidad algoritmica de cadenas cortas},pdfauthor={Hector Zenil}]{hyperref}
\hypersetup{colorlinks=true,
urlcolor=Cerulean,linkcolor=BrickRed,citecolor=RoyalBlue,a4paper,
  pdfpagemode=None,
  pdfstartview=FitH}
\usepackage[all]{hypcap}

\newenvironment{myenumerate}{
\begin{enumerate}
  \setlength{\itemsep}{1pt}
  \setlength{\parskip}{0pt}
  \setlength{\parsep}{0pt}}{\end{enumerate}
}

\begin{document}

\title{\textbf{Un m\'etodo estable para la evaluaci\'on\\de la complejidad algor\'itmica de\\cadenas cortas}\footnote{Para WCSCM2011. El autor H. Zenil no actu\'o como editor activo para este art\'iculo.}}\author{Hector Zenil\footnote{\href{mailto:h.zenil@sheffield.ac.uk}{\small{h.zenil@sheffield.ac.uk}}}\\ Dept. of Computer Science, University of Sheffield, Reino Unido\\ \medskip \& Centro de Ciencias de la Complejidad (C3), UNAM, M\'exico\\
Jean-Paul Delahaye\\Laboratoire d'Informatique Fondamentale de Lille (LIFL)\\Universit\'e de Lille 1, Francia}
\date{}
\maketitle

\begin{abstract}
Se discute y revisa un m\'etodo num\'erico propuesto que, de manera alternativa (y complementaria) al m\'etodo tradicional de compresi\'on, permite aproximar la complejidad algor\'itmica de cadenas, particularmente \'util para cadenas cortas para las cuales los m\'etodos tradicionales de compresi\'on no son efectivos y dependen de los algoritmos de compresi\'on. El m\'etodo muestra ser estable ya que produce clasificaciones razonables a partir de modelos de computaci\'on razonables, incluidos sistemas de etiquetas de Post, aut\'omatas celulares y m\'aquinas de Turing. Adem\'as, permite la concepci\'on (y comparaci\'on) de un modelo que predice la distribuci\'on de patrones en un mundo \medskip algor\'itmico.\\

\textbf{Palabras clave:} complejidad de Kolmogorov, probabilidad algor\'itmica de Solomonoff, teorema de codificaci\'on de Chaitin-Levin, semimedida de Levin, m\'aquinas de Turing peque\~nas, problema del castor atareado.
\end{abstract}

\section{Introducci\'on}

En el estudio de sistemas complejos, es fundamental contar no s\'olo con definiciones precisas sino tambi\'en con herramientas para evaluar la complejidad de sus objetos de estudio. El trabajo que hemos desarrollado y publicado en \cite{delahayezenil}, presenta una alternativa, confiable y estable \cite{pourlascience}, para evaluar la complejidad algor\'itmica (o de Kolmogorov) de cadenas de caracteres, en particular de cadenas cortas, para las cuales el m\'etodo tradicional de compresi\'on es, en la pr\'actica, in\'util para aproximar su complejidad algor\'itmica. La complejidad algor\'itmica de un objeto es la descripci\'on m\'as corta posible que regenera el objeto.

El m\'etodo introducido en \cite{delahayezenilcalude,delahayezenil} y recientemente difundido en la versi\'on francesa de la revista Scientific American \emph{Pour La Science} \cite{pourlascience}, prov\'ee un nuevo m\'etodo num\'erico y efectivo (hasta cierto punto) para la evaluaci\'on de la complejidad algor\'itmica de cadenas. En este art\'iculo lo describimos brevemente en el contexto de su relevancia como herramienta en el estudio e investigaci\'on en el \'area de sistemas complejos con su amplia y diversa gama de posibles aplicaciones debido a que, por primera vez, permite aproximar la complejidad de objetos que generalmente se utilizan en aplicaciones pr\'acticas, esto es, de objetos peque\~nos, cadenas de longitud corta. Por ejemplo en la compresi—n de datos, algoritmos de optimizaci—n, problemas de reconocimiento y clasificaci—n, por mencionar algunas. La mejor referencia introductoria al tema y aplicaciones es \cite{li}

\subsection{Pseudomedidas de complejidad}

Algunos m\'etodos atractivos para calcular la complejidad de cadenas cortas son insuficientes porque no coinciden con la complejidad de Kolmogorov de cadenas cuando aumenta la longitud que se requiere para que la medida sea consistente. Por ejemplo, una medida largamente pero erroneamente utilizada es la entrop\'ia de Shannon \cite{shannon} que se define como $-\displaystyle\sum p_i \log p_i$ (donde $p_i$ es una frecuencia). Con esta medida, la cadena 01010101010101010101 es la cadena con mayor entrop\'ia de Shannon posible para una cadena de longitud 20 (ya que tiene tantos ``0''s como ``1''s). Mientras que, la cadena 10010111010100001011,  tambi\'en de longitud 20, tiene la misma entrop\'ia de Shannon pero nuestra intuici\'on nos dice que deber\'ia ser m\'as compleja. 

La entrop\'ia de Shannon no hace mas que contar el n\'umero de ``0"s y ``1"s en una cadena ya que nunca fue dise\~nada para medir la complejidad (o el ``orden'') de la informaci\'on contenida en la cadena. Es una medida estad\'istica que ni siquiera es capaz de considerar las repeticiones y que hereda los problemas de la teor\'ia cl\'asica de probabilidades que, precisamente, la teor\'ia de la complejidad algor\'itmica (la complejidad de Kolmogorov) resuelve. De hecho, la entrop\'ia de Shannon es simplemente un corolario de la complejidad de Kolmogorov: si un objeto es m\'as complejo su transmisi\'on toma, potencialmente, m\'as tiempo. 

Otras medidas, como la complejidad por factores, por ejemplo, son erroneamente utilizadas cuando no hacen m\'as que cuantificar el n\'umero de posibles maneras de ordenar un sistema, son medidas probabilistas que en nada (o muy poco) se relacionan con una medida de complejidad. Si los investigadores en sistemas complejos est\'an interesados en una medida combinatoria (por ejemplo, el n\'umero de capas, de elementos en interacci\'on, etc.), con las limitaciones producto de las bases probabilistas en las que se funda (como es el caso, por mencionar otro ejemplo, del par\'ametro lambda de Langton \cite{langton}), los investigadores pueden continuar utilizando medidas \emph{ad-hoc} en el entendido de que no son una medida de complejidad universal (es decir, una medida de complejidad general y objetiva que pueda aplicarse en cualquier situaci\'on y a cualquier sistema).

Hoy en d\'ia, una amplia variedad de conceptos, basados en medidas como la entrop\'ia de Shannon (que Shannon dise\~no con el prop\'osito de cuantificar el ancho de banda necesario para un canal de comunicaci\'on) y otras falsas medidas de complejidad, se emplean para calcular y comparar la complejidad de objetos discretos bajo la falsa idea de que es el n\'umero de elementos o interacciones en un sistema hacen que un sistema sea complejo (por ejemplo, en la teor\'ia de sistemas din\'amicos, sistemas incluso muy simples resultan comportarse ca\'oticamente, incluso en sistemas de c\'omputo deterministas y extremadamente simples sin interacci\'on con el medio, producen aleatoriedad aparente y impredictabilidad \cite{wolfram}). Algunos autores podr\'ian argumentar que hay otras medidas de complejidad, pero medidas de complejidad como la profundidad l\'ogica de Bennett \cite{bennett}, est\'an fundadas, o bien son variaciones de la complejidad de Kolmogorov que toman en cuenta otros par\'ametros como el tiempo, la geometr\'ia de la evoluci\'on de un sistema o son versiones computables \cite{li} (que asumen recursos finitos) e interesantes de la complejidad de Kolmogorov, mientras que la mayor\'ia del resto de las pseudo medidas de complejidad est\'an fundadas en distribuciones de probabilidad o densidad, como la medida de Shannon o el par\'ametro de Langton.

\subsection{Complejidad algor\'itmica de Kolmogorov-Chaitin}

Imaginemos que se nos proporcionan dos cadenas cortas y se nos pregunta cu\'al de ellas parece ser el resultado de un proceso que genera cada s\'imbolo de la cadena al azar. Digamos que las cadenas son binarias y cortas, por ejemplo 0101 y 1011. A simple vista, la primera cadena tiene un patr\'on, aunque se repita s\'olo dos veces, y que podr\'ia ser aprovechado para generar una descripci\'on de la cadena. En espa\~nol, por ejemplo, la primera cadena podr\'ia ser descrita como ``dos veces cero y uno'' (aunque el lenguaje se presta a confusiones, ya que la misma descripci\'on puede interpretarse como 001 si no se conoce la longitud de la cadena\footnote{En ingl\'es, \'este tipo de ambig\"uedades regularmente se pueden evitar con la introducci\'on de una coma. As\'i ``zero, and one twice'' y ``zero and one twice'', engendran 011 y 0101 respectivamente. En espa\~nol, la soluci\'on es ``cero y uno dos veces'' versus ``cero y uno, dos veces'', pero en general la gram\'atica en espa\~nol (y otros idiomas, por ejemplo, franc\'es) no permite comas antes de la conjunci\'on `y' lo que no permite resolver todos los casos de ambig\"uedad de este tipo.}). Por otro lado, la segunda cadena parece necesariamente requerir una descripci\'on ligeramente m\'as larga. La primera podr\'ia describirse tambi\'en como ``cero seguido de uno seguido de cero seguido de uno''. Descripciones de la segunda pueden incluir ``uno y cero seguido de dos unos'' o ``uno, cero, uno, uno'', que no parece \'esta \'ultima una versi\'on comprimida de la cadena, sino m\'as bien una traducci\'on a una forma expandida del idioma. De hecho, pareciera que cadenas con patrones permiten menos descripciones distintas (int\'entese, por ejemplo, con cadenas m\'as largas).

Para resolver si alguna de las dos cadenas es, sin lugar a dudas, m\'as sencilla que la otra, o si la aparente repetici\'on de la primera cadena puede realmente aprovecharse a pesar de repetirse s\'olo dos veces, es necesario fijar un lenguaje objetivo (y que no permita las ambig\"uedades del lenguaje coloquial). Para determinar cu\'al de las cadenas parece m\'as aleatoria que la otra bastar\'ia, entonces, comparar sus respectivos valores de complejidad. La complejidad algor\'itmica de una cadena es el programa m\'as corto, medido en n\'umero de bits, que produce una cadena dada cuando se ejecuta en una m\'aquina universal de Turing. Asumimos que el lector est\'a familiarizado con el concepto de m\'aquina de Turing y de m\'aquina universal de Turing. Para una buena introducci\'on v\'ease \cite{minsky}.

El concepto de complejidad, introducido por Andrei Kolmogorov y Gregory Chaitin define la complejidad $K(s)$ de un objeto $s$ como el tama\~no del programa m\'as corto de computadora que genera $s$. Formalmente, $K_U(s) = \min\{|p|, U(p)=s\}$ donde $|p|$ es la longitud de $p$ medido en bits. 
En otras palabras, el tama\~no de un archivo comprimido $s$ es la complejidad de $s$. La complejidad de Kolmogorov (o Kolmogorov-Chaitin, para ser justos) proporciona una medida de \emph{aleatoriedad}.

La complejidad algor\'itmica es considerada la medida universal de complejidad. Sin embargo, no existe algoritmo efectivo que, para una cadena, el algoritmo produzca el entero que corresponda a la longitud del programa m\'as corto (la mejor compresi\'on posible) que genere la cadena como salida (el resultado se debe al problema de la detenci\'on de las m\'aquinas de Turing). Lo que significa que uno no puede medir con absoluta certeza la complejidad algor\'itmica de una cadena.

El que sea no computable no significa, sin embargo, que no se le pueda utilizar ya que en realidad a menudo se le puede aproximar de manera eficaz. El c\'alculo del valor aproximado de la complejidad de Kolmogorov, gracias a algoritmos de compresi\'on sin p\'erdida, hacen del concepto una herramienta de gran utilidad usado en diversas aplicaciones. De hecho existen aplicaciones de la teor\'ia de la complejidad algor\'itimca que han resuelto problemas de clasificac\'on de todo tipo de objetos \cite{li,zenilca}, para estudiar la similitud de ciertos idiomas, especies de animales, para detectar fraudes (por ejemplo, plagios) y caracterizar im\'agenes \cite{zenilgaucherel}.

\subsection{El problema de las cadenas cortas}

La complejidad de Kolmogorov permite una caracterizaci\'on matem\'atica del azar: una cadena aleatoria $s$ de $n$ bits de informaci\'on es una cadena cuya complejidad de Kolmogorov $K(s)$ es cercana a $n$. Es decir, el programa que lo produce es de m\'as o menos el mismo tama\~no en bits que la cadena original. Una cadena aleatoria infinita es tal que ning\'un proceso de compresi\'on puede comprimir por m\'as de una constante ning\'un segmento inicial de la cadena. Por ejemplo, la secuencia infinita 01010101... no es aleatoria, porque uno puede definir de forma concisa la ``serie infinita de 01'' y, sobre todo, escribir un programa muy corto basado en un bucle que genere la secuencia infinita. La secuencia compuesta de los d\'igitos de $\pi$ tampoco son aleatorios: hay un programa m\'as corto que genera todos sus decimales. 

La forma de abordar la complejidad algor\'itmica de una cadena es por medio del uso de algoritmos de compresi\'on sin p\'erdida. \emph{Sin p\'erdida} significa que se puede recuperar la cadena original a partir de la versi\'on comprimida por medio de un programa de descompresi\'on. Entre m\'as compresible se considera menos compleja la cadena. Por el contrario, si no es compresible, se le considera a la cadena como aleatoria o m\'aximamente compleja. El resultado de un algoritmo de compresi\'on es una cota superior de su complejidad algor\'itmica, por lo que se dice que la complejidad de Kolmogorov es computable por \emph{arriba}. Esto quiere decir que a pesar de que uno nunca puede decir cuando una cadena no es compresible, si uno tiene \'exito en la reducci\'on de la longitud de una cadena se puede decir que la complejidad algor\'itmica de esa cadena no puede ser mayor a la longitud de la versi\'on comprimida. 

Para evitar hacer trampa y decir que uno puede comprimir cualquier cadena con un algoritmo de compresi\'on \emph{ad hoc} (por ejemplo, codificando artificialmente ciertas cadenas complicadas con programas cortos interpretados en una m\'aquina universal truqueada) la codificaci\'on de la m\'aquina debe ser parte de la complejidad de un objeto cuando es medida con respecto a esa m\'aquina. Un algoritmo de compresi\'on transforma una cadena comprimida en dos partes: una parte es la versi\'on comprimida del objeto original, y la otra las instrucciones para descomprimir la cadena. Ambas partes deben ser contabilizadas en el tama\~no final de la cadena comprimida, debido a que se requieren las instrucciones de decompresi\'on para obtener la cadena original sin necesidad de depender de la elecci\'on arbitraria del algoritmo (o de la m\'aquina de Turing). En otras palabras, uno puede considerar que agrega el algoritmo de descompresi\'on a la cadena comprimida de manera que la cadena comprimida sea \emph{autodescomprimible} y venga con sus propias instrucciones de descompresi\'on\footnote{De hecho, algunos programas, como GZIP, permiten la generaci\'on de archivos comprimidos ejecutables, que empacan precisamente las instrucciones de descompresi\'on en el programa mismo y no requiere ni siquiera de tener instalado GZIP para descomprimirlo}. A la larga, un teorema \cite{li} (llamado de \emph{invarianza}) garantiza que los valores de la complejidad convergen a pesar de la elecci\'on arbitraria de lenguajes de programaci\'on o la utilizaci\'on de m\'aquinas de Turing truqueadas (en otras palabras, uno no puede seguir enga\~nando por siempre). 

El teorema de invarianza \cite{kolmogorov,chaitin} acota la diferencia entre evaluaciones de la complejidad de Kolmogorov calculadas con diferentes m\'aquinas de Turing. Si $U$ y $U^\prime$ son dos m\'aquinas de Turing universales diferentes, el teorema estipula que si $K_U(s)$ es la complejidad algor\'itmica de una cadena $s$ medida con respecto a una m\'aquina universal $U$ y $K_{U^\prime}(s)$ es la complejidad algor\'itmica de la misma cadena $s$ medida con respecto a otra m\'aquina universal $U^\prime$ entonces $|K_U(s) - K_{U^\prime}(s)|<c$ donde $c$ es una constante que no depende de $s$. En otras palabras, la diferencia en las evaluaciones es a lo m\'as la longitud finita de un compilador que pueda escribirse entre $U$ y $U^\prime$.

Uno requiere la utilizaci\'on de m\'aquinas universales porque es la \'unica manera de garantizar que la m\'aquina va a producir la cadena que elijamos evaluar y no se est\'e restringido a poder preguntarse sobre la complejidad de un conjunto limitado de cadenas (por ejemplo, cadenas producidas por lenguajes regulares).

El teorema de invarianza muestra que un sentido amplio la complejidad de Kolmogorov es una medida objetiva y universal. Aunque el teorema de invarianza le da estabilidad a la definici\'on de complejidad de Kolmogorov, tambi\'en hace evidente que, para cadenas cortas la medida es inestable porque la constante implicada ($c$), o sea el tama\~no de la m\'aquina universal (o las instrucciones de descompresi\'on) dominan el resultado final en la evaluaci\'on de $K(s)$. Es decir, incluir las instrucciones de descompresi\'on afecta la complejidad relativa de una cadena si la complejidad de la cadena es m\'as peque\~na que la longitud de las instrucciones de descompresi\'on, lo que resulta en evaluaciones inestables cuando se trata de cadenas cortas, es decir, cadenas de longitud cercana o menor a la longitud t\'ipica de las instrucciones de descompresi\'on (en el orden del tama\~no en bits del algoritmo de descompresi\'on).

Hasta ahora, a diferencia de cadenas suficientemente largas para las cuales los m\'etodo de compresi\'on funcionan, no exist\'ia un m\'etodo para evaluar la complejidad algor\'itmica de cadenas cortas, y por lo tanto una manera objetiva de determinar si una cadena como 000 es m\'as simple que 01101001 a pesar de que la intuici\'on nos sugiere que la primera parece m\'as simple y la segunda m\'as aleatoria.  Nos gustar\'ia decir objetivamente que, por ejemplo, la cadena de 7 bits 1001101 parece m\'as compleja que la cadena 0000000, o que 0001000 tiene una complejidad intermedia a las dos anteriores. As\'i que tenemos una idea intuitiva de una clasificaci\'on de complejidad pero no una medida objetiva que valide la intuici\'on. ?`C\'omo hacer que la teor\'ia confirme la intuici\'on, y que sea universal y consistente tanto para cadenas cortas como largas?

\subsection{El problema del m\'etodo de compresi\'on}

Para cadenas cortas (que son a menudo las usadas en aplicaciones pr\'acticas), la adici\'on de las instrucciones de descompresi\'on de la versi\'on comprimida hace que la cadena comprimida, con frecuencia, resulte m\'as larga que la versi\'on original. Si la cadena es, por ejemplo, m\'as corta que el tama\~no del algoritmo de descompresi\'on, no habr\'a forma de comprimir la cadena en algo m\'as corto que la longitud original de la cadena, simplemente porque las instrucciones de la descompresi\'on rebasan la longitud de la cadena original (Figura 1). Por otra parte, el resultado depende tanto del tama\~no del algoritmo de descompresi\'on (porque en estos casos es el mayor contribuyente a la longitud total) y por lo tanto la longitud (y aproximaci\'on de la complejidad algor\'itmica) es demasiado inestable. 

A manera de ilustraci\'on, si se trata de comprimir una cadena corta con, digamos, el lenguaje de programaci\'on \emph{Mathematica}, se obtiene que la longitud de la versi\'on comprimida de la cadena de longitud 0101 es: 
\begin{verbatim}
StringLength@Compress[``0101''] = 30
\end{verbatim}
(incluso antes de que la versi\'on comprimida sea transformada a bits para que el resultado est\'e en el mismo lenguaje de la cadena misma)

Esto significa que la compresi\'on de la cadena 0101 requiere de un programa de 46 caracteres (a\'un m\'as en bits) para ser generada, lo que no tiene sentido alguno, pues la simple descripci\'on en espa\~nol es m\'as corta que la versi\'on comprimida con \emph{Mathematica}. En \emph{Mathematica}, las cadenas comienzan a ser mejor comprimidas (en caracteres) cuando las cadenas tienen una longitud de 30 bits.  Si se trata de comprimir 1011 se llega nada menos que al mismo valor que para 0101, es decir: \begin{verbatim}
StringLength@Compress[``1011''] = 30
\end{verbatim}

\'Este no es, sin embargo, un fallo de \emph{Mathematica} sino el resultado de lo que hemos explicado. La funci\'on Compress en \emph{Mathematica} en realidad est\'a basada en el algoritmo de compresi\'on sin p\'erdida Deinflate, que es una combinaci\'on del algoritmo LZ77 y Huffman, dos de los algoritmos de compresi\'on sin p\'erdida m\'as populares disponible del mercado, utilizados en formatos p\'ublicos como ZIP, GZIP, GIF y PNG. 

\begin{figure}[htdp]
\centering
   \scalebox{.65}{\includegraphics{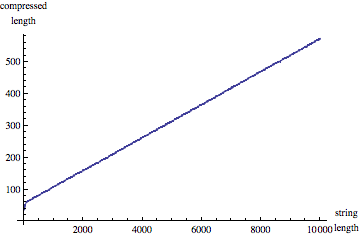}}
\caption{Gr\'afica de compresi\'on de una cadena de $n$ bits (eje $x$) contra su versi\'on comprimida (eje $y$) usando un t\'ipico algoritmo de compresi\'on de datos. Al principio de la l\'inea de compresi\'on se observa que el origen no pasa por $y = 0$, incluso cuando $x = 0$, lo que significa que cadenas cortas comprimidas resultan m\'as largas que su tama\~no original.}
\end{figure}

\begin{figure}[htdp]
\centering
   \scalebox{.65}{\includegraphics{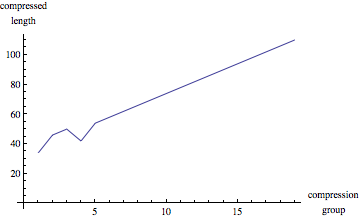}}
\caption{Al hacer un acercamiento al origen de esta gr\'afica de compresi\'on de una cadena de $n$ ``1''s (eje $x$), contra las diferentes longitudes de sus versiones comprimidas (eje $y$), se verifica que el inicio es m\'as inestable que el resto.}
\end{figure}

Las instrucciones obviamente ocupan un espacio del valor final de la longitud comprimida y no pueden ser ellos mismos (las instrucciones) comprimidas (si lo fueran, ser\'ia en todo caso una longitud constante para todas las cadenas, que nos remiten a la misma situaci\'on). En resumen, hay un l\'imite para los algoritmos de compresi\'on para comprimir cadenas cortas. As\'i que si se quisiera decir cu\'al de las dos cadenas son objetivamente m\'as o menos complejas por medio de la aproximaci\'on de su complejidad algor\'itmica mediante un algoritmo de compresi\'on, resulta que no hay manera de obtener una respuesta, por el contrario, se encuentra una medida inestable y generalmente sin sentido (Figura 2).

\subsubsection{El problema del bit isolado}

Por ejemplo, dada la definici\'on de la complejidad algor\'itmica basada en la compresibilidad, si una cadena no es compresible entonces es aleatoria, de donde de inmediato se podr\'ia decir que un bit isolado, 0 \'o 1, son cadenas al azar con toda certitud, ya que tienen complejidad algor\'itmica m\'axima, dado que no hay manera de comprimir un solo bit en algo m\'as corto (siendo el bit indisociable, la unidad m\'inima y b\'asica). En otras palabras, no hay un programa de menos de 1 bit que produzca 0 o 1. La mejor descripci\'on de 0 y 1 son, por lo tanto, 0 y 1 mismos. Por lo tanto, la teor\'ia dice que son autom\'aticamente cadenas aleatorias. Evidentemente, esto puede chocar con nuestra intuici\'on de azar si se les compara con cadenas m\'as largas y que parecen m\'as el resultado de azar, o al menos se les puede considerar m\'as complejas (por ejemplo, 0110101).

Por un lado, un bit solo no contiene informaci\'on alguna, y por este motivo uno podr\'ia pensar que representa de alguna manera al azar. Si uno piensa si uno habr\'ia sido capaz de predecir 0 o 1 como el resultado de un proceso, dado que no hay ning\'un contexto, ya que se producen solos, se podr\'ia concluir que su ocurrencia es de alguna manera el resultado (aparente o no) del azar. En otras palabras, si uno ve una cadena como 010101, uno apostar\'ia f\'acilmente que el siguiente bit es un 0, pero si no se le proporciona mas que un bit no hay manera de favorecer uno u otro resultado siguiente.

Es dif\'icil, sin embargo, justificar c\'omo la cadena de un solo bit ``0" pod\'ia parecer m\'as aleatoria que, digamos, cualquier otra cadena posible, sino es bajo el razonamiento descrito anterior, que se refiere a contextos y no la noci\'on de compresibilidad. La intuici\'on nos dice que las cadenas cortas (incluido ``0" o ``1") no parecen m\'as aleatorias que cualquier otra cadenas posible, y que si un bit representa la m\'axima complejidad entre todas las cadenas finitas, y otras cadenas cortas como 000 no son aleatorias, entonces hay una fase de transici\'on particularmente abrupta entre las cadenas de 1 bit y cadenas de unos cuantos bits m\'as, lo que parece contraintuitivo.

El problema a resolver es, como Delahaye ha se\~nalado en \cite{pourlascience}, un problema de term\'ometro: entre todos los instrumentos de medici\'on que conducen a la evaluaci\'on de la complejidad de Kolmogorov, pero que difieren por constantes aditivas ?`cu\'al es el mejor? ?`c\'omo elegir? Una soluci\'on es utilizar el m\'etodo que hemos dise\~nado. En lugar de elegir un s\'olo termometro, o una sola medida, enumeramos todas y nos fijamos en un conjunto suficientemente grande de ellas. El m\'etodo le va a dar sentido tambi\'en al problema del bit isolado.

\section{La probabilidad algor\'itmica de Solomonoff-Levin}

La intuici\'on nos dice que algo aleatorio tambi\'en debe ser raro y poco com\'un. Si uno se pregunta qu\'e tan com\'un es 0 o 1 como resultado de la ejecuci\'on de un programa elegido al azar, hay una medida que indica la probabilidad de que un programa produzca una cadena determinada si se ejecuta en una m\'aquina universal de Turing. \'Esta es la medida que utilizamos para presentar un nuevo m\'etodo para evaluar la complejidad algor\'itmica de cadenas, incluyendo cadenas cortas incluso de un bit, como alternativa al uso tradicional de los algoritmos de compresi\'on. El nuevo m\'etodo tiene como objetivo para resolver el problema de la evaluaci\'on de la complejidad de las cadenas cortas, como hemos discutido; y resuelve el problema del bit isolado. Se basa en el concepto de probabilidad algor\'itmica de Solomonoff -Levin y se conecta con la complejidad algor\'itmica (Kolmogorov-Chaitin) por medio del teorema de codificaci\'on de Chaitin-Levin. 

Este nuevo m\'etodo que resuelve varios problemas que la teor\'ia y los m\'etodos actuales no permit\'ian resolver, tiene una limitaci\'on: es muy costoso en t\'erminos de computaci\'on. Como la duraci\'on o las longitudes muy cortas, los objetos de complejidad muy d\'ebil son muy dif\'iciles de evaluar y, parad\'ojicamente, el m\'etodo de evaluaci\'on requiere de un c\'alculo masivo. En la pr\'actica, s\'olo proporciona resultados para cadenas muy cortas y desde este punto de vista los m\'etodos de compresi\'on siguen siendo esenciales para complementar la necesidad de aproximar la complejidad algor\'itmica de cadenas largas. Delahaye hace una analog\'ia interesante: Al igual que en astronom\'ia, donde, dependiendo del tipo de objetos y la distancia, se utiliza uno u otro m\'etodo para calcular distancias, en las medidas de complejidad, nuestro m\'etodo proporciona una alternativa para cadenas cortas y se pueden adoptar m\'etodos h\'ibridos con la utilizaci\'on de algorithmos de compresi\'on y t\'ecnicas de concatenaci\'on (en particular para cadenas de tama\~no mediano).

La idea que subyace nuestro nuevo term\'ometro de baja complejidad para cadenas cortas se basa en una propiedad notable de la complejidad de Kolmogorov: entre m\'as un objeto es simple, m\'as se produce con frecuencia cuando se utiliza una computadora ejecutando programas al azar.

En un lenguaje de programaci\'on Turing completo (es decir, en el que cualquier funci\'on computable puede implementarse) si cada secuencia de instrucciones es generada de manera aleatoria y ejecutada, muy frecuentemente producir\'a un programa gramaticalmente inv\'alido que no puede si quiera ejecutarse. Otras muchas veces el programa va a comenzar a ejecutarse y no se detendr\'a jam\'as. Estos programas no nos interesan, s\'olo nos interesan aquellos que producen como salida una cadena finita de ``0''s y ``1''s y se detienen. 

Evidentemente si se ejecutan varios programas distintos algunos van a producir la misma salida (para la misma entrada). Si ejecutamos tantos programas como podamos podemos generar una clasificaci\'on de frecuencia en donde a cada cadena se le asigna una frecuencia de repetici\'on $r$ de entre todos los programas $t$ ejecutados. Esto define una distribuci\'on de probabilidad sobre $\{0,1\}^n$, es decir, sobre todas las cadenas binarias.

Como resultado, obtenemos que cadenas como 0000000 son mucho m\'as frecuentes que cadenas como 1001101. El resultado es una distribuci\'on de potencia en donde las cadenas m\'as frecuentes tienen baja complejidad y las menos frecuentes mayor complejidad (son m\'as aleatorias). Esta probabilidad es la medida que Solomonoff \cite{solomonoff} y Levin \cite{levin} caracterizaron matem\'aticamente mediante un razonamiento relativamente sencillo (tirar programas al azar). En resumen, para Kolmogorov, la complejidad de una cadena es una longitud mientras que para Solomonoff es una probabilidad.

Formalmente, si $m(s)$ es la probabilidad de producci\'on de $s$, $m(s) = \displaystyle\sum_{p : U(p) = s} 2^{-|p|} = pr(U(p) = s)$. Es decir, la suma sobre todos programas $p$ que al ejecutarse en una m\'aquina (autodelimitada) universal de Turing $U$ generan $s$ y se detienen, o si se prefiere, la probabilidad de que $U$ corriendo $p$ produzca $s$. 

Una m\'aquina de Turing autodelimitada o prefix-free es una m\'aquina cuyas entradas forman una codificaci\'on prefix-free, esto quiere decir que ninguna entrada es el principio de ninguna otra. Esto es para garantizar ciertas propiedades de la medida por varias razones. M\'as detalles pueden encontrarse en \cite{calude,li}. Una codificaci\'on prefix-free es, por ejemplo, el sistema mundial de n\'umeros telef\'onicos. Si un n\'umero telef\'onico fuera el principio de uno otro, nunca le ser\'ia posible a uno comunicarse con el del n\'umero de tel\'efono m\'as largo pues como funciona la red de telefon\'ia mundial, la secuencia de n\'umeros que formen un n\'umero de tel\'efono valido es inmediatamente utilizado para realizar la conexi\'on. Imagina por un momento que mi n\'umero de tel\'efono fuera 558973213 y que el de otra persona fuera 558973. Evidentemente, el que intente marcar mi n\'umero siempre acabar\'a comunic\'andose con 558973. Una codificaci\'on prefix-free permite poder hablar de la probabilidad de un conjunto de programas que produzcan cierta cadena sin que el conjunto pueda acotarse de alguna forma.

\subsection{El teorema de codificaci\'on de Chaitin-Levin}

Los valores de $m$ est\'an relacionados con la complejidad algor\'itmica porque el t\'ermino m\'as grande en la sumatoria es el programa m\'as corto, y por tanto, es $K(s)$ quien domina el total. Un teorema, clave para el m\'etodo que propusimos para evaluar la complejidad de Kolmogorov, relaciona matem\'aticamente $m(s)$ y $K(s)$. El teorema de codificaci\'on de Chaitin-Levin establece que $K(s)$ es aproximadamente igual a $-\log_2(m(s))$. En otras palabras, $K(s)$ y $m(s)$ difieren de una constante c, independiente de la cadena $s$ tal que $|-\log_2($m(s)$)-K(s)|<c$. 

A groso modo, la probabilidad algor\'itmica $m$ dice que si hay muchas descripciones largas de cierta cadena, entonces tambi\'en hay una descripci\'on corta y por lo tanto con baja complejidad algor\'itmica  y si hay pocas descripciones para una cadena, entonces dif\'icilmente tendr\'a una descripci\'on corta.

Que el logaritmo negativo de $m(s)$ coincida con la complejidad algor\'itmica de $s$ con una diferencia de una constante significa que aproximar $m(s)$ nos aproxima a $K(s)$. Debido a que ni  $K(s)$ ni $m(s)$ son computables, no hay programa que tome $s$ como entrada y produzca $m$, $m$ tiene que ser tambi\'en aproximado en lugar de calculado con certeza absoluta.

\section{C\'alculo de la probabilidad de producci\'on}

El c\'alculo de $m(s)$ se obtiene mediante la ejecuci\'on de un gran n\'umero de programas que ser\'an producidos al azar, o de un conjunto de programas que se enumeran de manera sistem\'atica (que resulta en lo mismo). Al combinar los resultados te\'oricos discutidos anteriormente se obtiene una distribuci\'on que llamaremos $SL(s)$, y de donde podremos calcular $-\log_2(SL(s))$ para aproximar $K(s)$. $SL$ es una distribuci\'on que puede escribirse en funci\'on del n\'umero de estados de las m\'aquinas que se utilizan para generar la distribuci\'on de frecuencia de salida de las m\'aquinas de Turing. Como funci\'on, $SL$ no es computable ya que si lo fuera, es decir si se pudiera calcular num\'ericamente $SL$ para cualquier n\'umero de estados de m\'aquinas de Turing, se podr\'ia resolver el problema del castor atareado para cualquier n\'umero de estados, lo que se sabe es imposible por contradicci\'on con el resultado de Rado de incomputabilidad de las funciones del castor ateareado.

\subsection{M\'aquinas de Turing peque\~nas}

Usamos un m\'etodo de c\'alculo tan primitivo como posible, pero lo suficientemente poderoso para que cualquier programa pueda ser ejecutado y cualquier cadena producida. Introducido por Alan Turing en 1936, el modelo de las m\'aquinas de Turing ha desempe\~nado un papel fundamental en la ciencias de la computaci\'on y la l\'ogica matem\'atica, ya que ha permitido el estudio de lo que es un algoritmo y, en la pr\'actica, del desarrollo de la computadora digital. El modelo de Turing puede verse como un lenguaje de programaci\'on; la descripci\'on de una m\'aquina de Turing es equivalente a escribir un programa de c\'omputaci\'on. 

Las m\'aquinas de Turing son el modelo de computaci\'on m\'as conocido, debido a que es un modelo que tiene una representaci\'on f\'isica cuya motivaci\'on fue la descripci\'on de un humano que calculara con l\'apiz y papel. Podemos verlas como una abstracci\'on de nuestras computadoras. Disponen de una cinta de longitud ilimitada dividida en celdas discretas (an\'aloga a la tira de papel donde escrib\'ia el computador humano) sobre la que se sit\'ua una cabeza capaz de leer y escribir en la celda donde se encuentra. La m\'aquina s\'olo lee y escribe un conjunto finito de s\'imbolos conocido como su alfabeto. Entre estos s\'imbolos hay uno llamado usualmente blanco que es el que por defecto llena todas las celdas de la cinta. Existe un conjunto finito de estados en los que puede encontrarse la m\'aquina. Uno de tales estados es el estado inicial desde el que comienzan todas las computaciones. Tambi\'en suele haber un estado de detenci\'on, que cuando se alcanza se termina la computaci\'on. En cada paso de computaci\'on, la m\'aquina de Turing:

\begin{myenumerate}
\item Lee el s\'imbolo escrito en la celda sobre la que se encuentra la cabeza. 
\item En funci\'on del s\'imbolo le\'ido y del estado actual de la m\'aquina:
\begin{myenumerate}
\item Escribe un nuevo s\'imbolo en la celda (puede ser igual al que hab\'ia). 
\item Se desplaza una posici\'on a la izquierda o derecha sobre la cinta. 
\item Cambia de estado (o permanece en el mismo).
\end{myenumerate}
\end{myenumerate}

As\'i se contin\'ua hasta llegar al estado de parada. Lo que caracteriza las computaciones de una m\'aquina de Turing es su tabla de transiciones. Si vemos la enumeraci\'on anterior, el comportamiento en cada paso de computaci\'on depender\'a del estado en que se encuentra la m\'aquina de Turing y el s\'imbolo le\'ido.

Las m\'aquinas de Turing constituyen el ejemplo m\'as conocido de dispositivo de computaci\'on abstracto capaz de computaci\'on universal, lo que significa que para cualquier funci\'on efectivamente calculable existe una m\'aquina de Turing que la calcula. Especialmente interesantes son las m\'aquinas de Turing universales, capaces de simular la computaci\'on de cualquier otra m\'aquina de Turing. 

El n\'umero de estados de una m\'aquina de Turing determina su poder de c\'alculo. M\'aquinas con un estado s\'olo pueden hacer c\'alculos sencillos, tales como invertir 0 a 1 y 1 a 0 a una cadena que se les presenten. Una m\'aquina que dispone de 2 estados comienza a hacer cosas m\'as interesantes. El n\'umero de m\'aquinas con 2 estados es, curiosamente, 10\,000 que por supuesto s\'olo pueden generar cadenas muy cortas (no m\'as largas que el m\'aximo n\'umero de pasos que una m\'aquina que se detiene puede alcanzar y que es acotado por el n\'umero de estados). Al operar todas las m\'aquinas de Turing de 3 estados, el n\'umero de m\'aquinas comienza a crecer de manera colosal pero \'estas generan cadenas m\'as largas que permiten el c\'alculo de frequencia, y por tanto la probabilidad de producci\'on de cadenas un poco m\'as largas.

\subsubsection{El problema de la detenci\'on}

Sin embargo, las m\'aquinas de Turing pueden detenerse o no dependiendo si entran en el estado de detenci\'on de su tabla de instrucciones (y del contenido de la cinta, que en este caso es siempre blanco). Aunque Alan Turing demuestra la existencia de una m\'aquina de Turing universal, es decir, una m\'aquina de Turing capaz de simular cualquier otra m\'aquina de Turing, tambi\'en muestra que no existe una m\'aquina de Turing que pueda determinar si cualquier otra m\'aquina se detendr\'a. A este problema se le conoce como el problema de la detenci\'on.

Evidentemente, si uno est\'a interesado en la salida de una m\'aquina de Turing, definida como el resultado de lo que contiene su cinta una vez que se detiene. Si no es posible saber si una m\'aquina se va a detener no hay manera de determinar con certeza su salida, ni la frecuencia de una cadena en general.

Una manera elegante y concisa de representar el problema de la detenci\'on es el n\'umero de Chaitin $\Omega$ \cite{chaitin} (un n\'umero irracional entre 0 y 1), cuyos d\'igitos en su expansi\'on binaria es la probabilidad de detenci\'on de una m\'aquina de Turing universal corriendo programas al azar. Formalmente, $0 < \Omega = \displaystyle\sum_{\normalsize{p\textbf{ }se\textbf{ }detiene}} 2^{-|p|} < 1$ con $|p|$ la longitud de $p$ en bits\footnote{La definici\'on precisa requiere que la m\'aquina de Turing universal sea \emph{prefix-free} (para mayor informaci\'on v\'ease \cite{calude})}.

\subsection{El problema del castor atareado}
\label{beaver}

De entre las m\'aquinas que se detienen una pregunta, realizada por Rado \cite{rado}, es cu\'al m\'aquina de $n$ estados (y 2 s\'imbolos) escribe m\'as s\'imbolos o le toma m\'as tiempo para detenerse a partir de una cinta en blanco. Al m\'aximo n\'umero de pasos se le asigna un n\'umero $S(n)$ que depende solamente del n\'umero de estados $n$ y se le llama a dicha m\'aquina un \emph{castor atareado} (o \emph{busy beaver} en ingl\'es) com\'unmente denotado por $B(n)$.

Ahora bien, si se conoce el valor $S(n)$ para $B(n)$ cualquier m\'aquina que corra m\'as de $S(n)$ es una m\'aquina que no se detendr\'a nunca. As\'i que basta ejecutar cada m\'aquina para saber si se detiene o no. Rado demuestra, sin embargo, que la funci\'on $n \rightarrow S(n)$ no es computable, es decir, no existe un algoritmo (o m\'aquina de Turing) que dado un n\'umero de estados produzca el n\'umero $S(n)$.

El n\'umero $\Omega$ de Chaitin, el castor atareado, la probabilidad algor\'itmica y nuestro m\'etodo, est\'an todos \'intimamente relacionados. Para m\'aquinas de Turing peque\~nas, el problema de la detenci\'on se puede resolver porque, por un lado, porque no son relativamente muchas y uno puede ya sea ejecutar todas las m\'aquinas y examinar su comportamiento o examinar la tabla de instrucciones de la m\'aquina y decidir anal\'iticamente si se detiene o no. Sin embargo, la secuencia de n\'umeros del castor atareado, $S(1)$, $S(2)$, $\ldots$ crece m\'as r\'apido que cualquier secuencia computable. Porque si una m\'aquina de Turing pudiese computar una sucesi\'on que crece m\'as r\'apido que el castor atareado, entonces dicha secuencia, parad\'ojicamente, resolver\'ia el problema del castor atareado.

Es f\'acil verificar que para $B(1)$, $S(1)=1$ pues no hay mucho lugar para cualquier otro comportamiento m\'as complicado. Con dos estados, Rado determina que $S(2)=6$ y unos a\~nos despu\'es, junto con Lin \cite{lin}, probaron que $S(3)=21$ requiriendo un an\'alisis exhaustivo y un importante poder computacional. Brady \cite{brady}, usando t\'ecnicas de an\'alisis m\'as sofisticadas y a\'un un mayor poder computacional prueba que que $S(4)=107$, pero el valor de $S(5)$ es desconocido, aunque se conocen algunas cotas.

Un programa que muestra los valores de los castores atareados y su evoluci\'on est\'a disponible en l\'inea \cite{busyhz}.

\section{Evaluando la complejidad de cadenas cortas}

El hecho de conocer los valores del castor atareado permite acotar el c\'alculo sistem\'atico y masivo de un gran n\'umero de m\'aquinas de Turing para producir una clasificaci\'on de frecuencia de cadenas binarias. Esta consideraci\'on es, por supuesto, esencial para no perder tiempo innecesariamente en funcionamiento de m\'aquinas que no contribuyen a los resultados deseados. Para m\'aquinas de Turing con 3 estados, por ejemplo, cualquier m\'aquina que ejecute m\'as de 22 pasos, es una m\'aquina que no se dentendr\'a nunca, ya que para $B(3)$, $S(3)$=21. Para 4 estados $S(4)=107$, pero no se conoce $S(n)$ para $n>4$ y por lo tanto nuestro experimento exhaustivo s\'olo puede realizarse a lo m\'as para todas las m\'aquinas de 4 estados.

Para las  7\,529\,526  m\'aquinas de Turing con 2 s\'imbolos (0 y 1) y 3 estados, los resultados de $SL(s)$ comienzan a arrojar indicios de un ordenamiento no trivial de una clasificaci\'on de complejidad para cadenas binarias. Por ejemplo, entre las cadenas binarias de longitud 6, el c\'alculo de $SL(s)$ produce, mediante la ejecuci\'on de m\'aquinas de Turing con 3 estados, las cadenas 000000 y 111111 con la m\'as alta probabilidad (y por lo tanto la m\'as baja complejidad algor\'itmica), que es lo que uno podr\'ia esperar. Seguido en orden de las siguientes cadenas: 000001, 100000, 111110 y 011111 con igual frecuencia, seguidas de 000100, 001000, 111011 y 110111 con igual frecuencia, seguidas de 001001, 100100, 110110 y 011011, 010110 y finalmente por el conjunto 101001, 100101 y 011010. Esta clasificaci\'on es bastante sutil, pero natural para colocar la cadena que puede describirse como un 1 detr\'as de cinco 0 como m\'as simple que un 1 con tres 0 delante y dos detr\'as. La clasificaci\'on obtenida con la m\'aquinas de Turing tiene 3 estados producen s\'olo 128 diferentes cadenas con las cuales comparar. Esto deja a las dem\'as cadenas un poco m\'as largas, pero a\'un cortas, sin probabilidad. Por lo tanto, ejecutamos las m\'aquinas de Turing con 4 estados para obtener un mayor n\'umero de cadenas y generar una clasificaci\'on m\'as completa y fidedigna. Para ello fue necesario correr 11\,019\,960\,576 m\'aquinas de Turing, que a pesar de ciertos m\'etodos para acortar su c\'alculo llevo casi 9 d\'ias (en una sola computadora portatil con un procesador Duo a 1.2 Ghz. y 4Gb de memoria RAM) usando un programa escrito en lenguaje C, mediante la libreria \emph{bignum} para grandes n\'umeros ya que las tablas de transici\'on de cada m\'aquina se generaban en tiempo real a partir de una enumeraci\'on, ya que generar las reglas de manera combinatoria y almacenarlas resultaba, para este n\'umero de m\'aquinas, imposible para cualquier disco duro actualmente en el mercado (sin mencionar el tiempo de lectura que para cada m\'aquina tomar\'ia). Algunas simetr\'ias pudieron ser explotadas (por ejemplo, para toda regla de una m\'aquina de Turing existe una que es su complemento y basta calcular el complemento de su salida para conocer el resultado antes de correrla) reduciendo el tiempo de c\'omputo a los 9 d\'ias mencionados.

Las m\'aquinas de Turing con 4 estados producen 1824 diferentes cadenas binarias que permiten la aproximaci\'on de $K(s)$ a trav\'es de $SL(s)$ y la f\'ormula $-\log_2(SL(s))$. Hay que tener en cuenta que la complejidad de Kolmogorov calculada a trav\'es de $SL$ es un n\'umero real. Esto es realmente una ventaja ya que permite una clasificaci\'on m\'as fina, pero si se quiere interpretar el resultado como la longitud de un programa basta tomar el siguiente entero. 

As\'i que valores exactos pueden ser num\'ericamente aproximados mediante el uso delos valores conocidos del castor atareado hasta $n=4$ y hemos publicado las tablas completas en l\'inea \cite{algorithmicnature} y tablas parciales en \cite{delahayezenil}.

\subsection{Un m\'etodo estable, aut\'omatas celulares y un modelo de distribuci\'on de patrones}

Una pregunta fundamental, y evidente, es qu\'e tan estable y robusto es el m\'etodo si se utilizan diferentes formalismos de c\'omputo (por ejemplo, usando m\'aquinas de Turing con una cinta ilimitada en una sola direcci\'on o utilizando automatas celulares en lugar de m\'aquinas de Turing). Hemos mostrado que formalismos de c\'omputo razonables producen clasificaciones de complejidad razonables \cite{zenildelahaye}. Las mismas distribuciones de frecuencia fueron producidas ejecutando y explorando un espacio representativo de automatas celulares unidimensionales con 2 colores y rango $3/2$, es decir, el espacio de aut\'omatas celulares que utilizan en sus reglas de producci\'on el estado de 2 celdas a la izquierda y una a la derecha, y con condici\'on inicial la m\'as simple posible (una celda negra). Este espacio de aut\'omatas celulares, que es justo en tama\~no el espacio siguiente m\'as grande al de aut\'omatas celulares elementales \cite{wolfram} (es decir, unidimensionales, con rango 1 y 2 colores posibles) nos permiti\'o hacer una exploraci\'on del tipo de clasificaciones de complejidad de cadenas que producen. Se eligi\'o este espacio porque el espacio siguiente m\'as simple es el de los aut\'omatas celulares elementales definidos por Wolfram \cite{wolfram} que no contiene mas que 256 aut\'omatas celulares y por lo tanto un n\'umero no muy significativo. Evidentemente tanto para aut\'omatas celulares, como m\'aquinas de Turing o cualquier otro formalismo de computaci\'on, entre m\'as n\'umero de automatas explorados mejor. Sin embargo, las restricciones en tiempo y recursos de c\'omputo no permiten obtener resultados en un tiempo razonable, ni a\~naden necesariamente mayor informaci\'on al modelo descrito.

\begin{figure}
\label{comps}
\centering
   \scalebox{.6}{\includegraphics{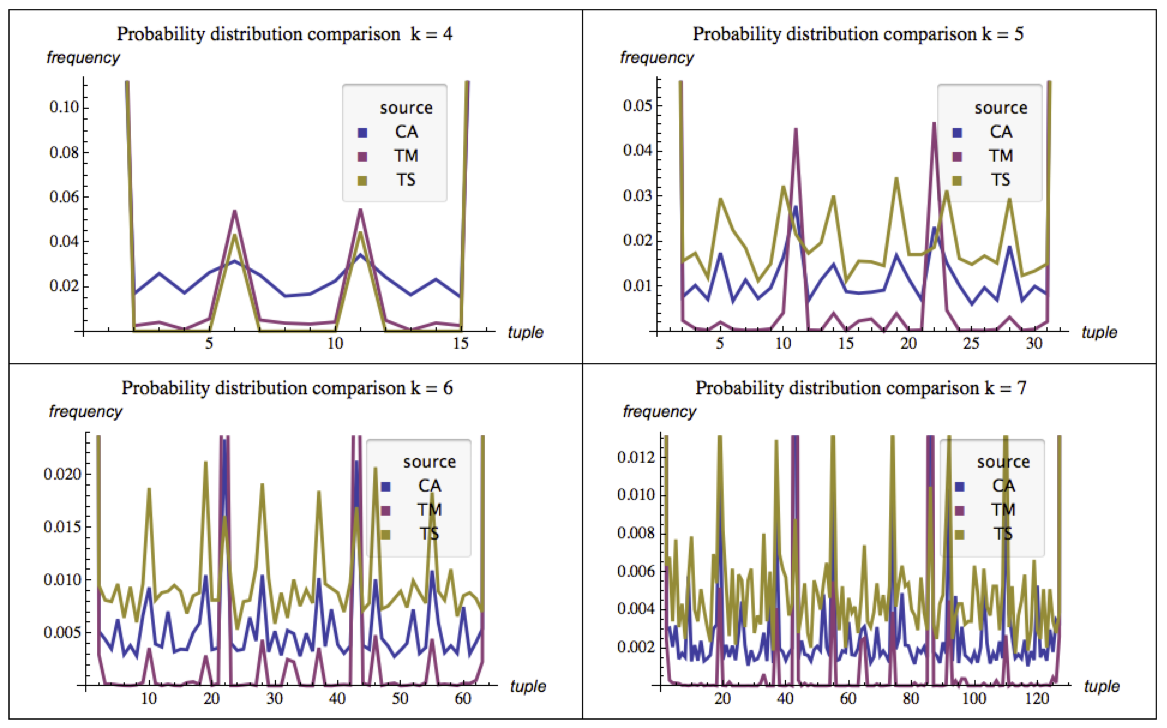}}
\caption{Comparaci\'on de $k$-tuplas generadas por aut\'omatas celulares (CA), m\'aquinas de Turing (TM) y sistemas de etiquetado de Post (TS). Las tuplas est\'an ordenadas por orden lexicogr\'afico.}
\end{figure}

A diferencia de las m\'aquinas de Turing, los aut\'omatas celulares favorecen ciertas simetr\'ias por la manera en que los aut\'omatas celulares evolucionan aplicando su regla en paralelo sobre todas las celdas al mismo tiempo. Adem\'as, tambi\'en a diferencia de las m\'aquinas de Turing, los aut\'omatas celulares no tienen un estado de detenci\'on (las cadenas que producen no contienen, por lo tanto, la \emph{informaci\'on} de su los l\'imites extremos que contiene una cadena producida por una m\'aquina de Turing que se detiene) ya que un automata celular se detiene en un tiempo arbitrario decidido por el que lo ejecuta (en nuestro caso, cada automata celular fue detenido arbitrariamente despu\'es de 100 pasos). Sin embargo, las clasificaciones producidas mediante aut\'omatas celulares (y tambi\'en sistemas de etiqueta de Post) resultaron parecerse unas a otras, la similitud fue cuantificada estad\'isticamente mediante el coeficiente de correlaci\'on de Spearman \cite{spearman} (coeficiente dise\~nado para comparar clasificaciones). El hecho de que las clasificaciones no s\'olo sean razonables con respecto a nuestra intiuici\'on de lo que es complejo o simple (por ejemplo, las cadenas 000... y 010101... aparecen con baja complejidad aleatoria mientras que cadenas que parecen aleatorias intuitivamente, lo son tambi\'en en las clasificaciones), sino adem\'as est\'en correlacionadas estad\'isticamente (Figura \ref{comps}) y sean compatibles con la definici\'on universal de complejidad algor\'itmica, proporcionan un m\'etodo eficaz, general y estable (las tablas con las clasificaciones completas est\'an disponibles en \url{http://www.algorithmicnature.org}).

El m\'etodo descrito tiene, por un lado, la remarcable caracter\'istica de resolver un problema te\'orico (el de la estabilidad de la definici\'on y evaluaci\'on de la complejidad de cadenas cortas, por ejemplo, el problema del bit isolado) y permite, en la pr\'actica y gracias a su estabilidad y consistencia, la comparaci\'on de la complejidad de distintas clasificaciones. Esta \'ultima ventaja permite, por ejemplo, comparar la distribuci\'on de patrones (cadenas con cierta complejidad) presente en el mundo f\'isico (a partir de fuentes de informaci\'on emp\'irica) y la clasificaci\'on producida por medios algor\'itmicos (utilizando las m\'aquinas de Turing o alg\'un otro formalismo). Sus similitudes y diferencias podr\'ian decirnos qu\'e tanto las estructuras que se forman en el mundo real pueden ser el resultado de procesos algor\'itmicos a diferencia de, por ejemplo, procesos mayoritariamente aleatorios. En \cite{zenildelahaye} y \cite{fqxi} nos hemos formulado estas preguntas, y esbozado un inicio de ruta de investigaci\'on.

\section{Comentarios finales}

El avance del programa de investigaci\'on que se describe aqu\'i pone a disposici\'on un m\'etodo general para calcular la complejidad de Kolmogorov. 

En el art\'iculo de \emph{Pour La Science} \cite{pourlascience} Delahaye se\~nala:

\begin{quote}
Pour les petites s\'equences, cette mesure est stable et conforme ˆ notre id\'ee de la complexit\'e, et, pour les grandes, elle est, d'aprs le th\'eorme mentionn\'e conforme ˆ la mesure de meilleure mesure de complexit\'e unanimement admise, la complexit\'e de Kolmogorov. Que demander de plus?

(Para cadenas cortas, esta medida [la medida que describo en este art\'iculo, nuestro comentario] es estable y se ajusta a nuestra idea de la complejidad, y, para largas cadenas, de acuerdo con el teorema mencionado [el teorema de invarianza, nuestro comentario], se ajusta a la mejor y universalmente aceptada medida de la complejidad, la complejidad de Kolmogorov. ?`Qu\'e m\'as se puede pedir?)
\end{quote}

El m\'etodo pretende ser utilizado para evaluar la complejidad de cadenas m\'as largas mediante su descomposici\'on en subcadenas m\'as cortas para las cuales podemos calcular su complejidad y generar una aproximaci\'on de la complejidad de la cadena original. Tampoco es necesario recorrer espacios completos para aproximar un valor de complejidad. Muestreos del espacio de m\'aquinas de Turing con 5 estados, espacios de aut\'omatas celulares unidimensionales con rangos de vecindad m\'as grandes y otros formalismos de computaci\'on, como sistemas de substituci\'on, pueden utilizarse. De hecho una pregunta abierta, es qu\'e tanto peque\~nas diferencias en un mismo formalismo impactan las medidas de complejidad. Por ejemplo, si se les permite a las m\'aquinas de Turing quedarse en la misma celda o no moverse m\'as que en una direcci\'on de la cinta (variantes que preservan universalidad).

A manera de conclusi\'on, Chaitin ha expresado \cite{chaitintesis} que (hablando de los resultados de nuestro m\'etodo): 

\begin{quote}
$\ldots$the dreaded theoretical hole in the foundations of algorithmic complexity turns out, in practice, not to be as serious as was previously assumed.

($\ldots$el agujero te\'orico terrible en los fundamentos de la complejidad algor\'itmica resulta, en la pr\'actica,  no ser tan grave como se supon\'ia anteriormente).
\end{quote}

Sin embargo, lo cierto es que estamos muy lejos de haber sacado todas las conclusiones y aplicaciones posibles.

\section*{Agradecimientos}

Los autores agradecen a los dos revisores cuyos comentarios y sugerencias fueron de gran valor para mejorar la manera de comunicar los resultados explorados en este art\'iculo. Cualquier error en \'el, sin embargo, es exclusiva responsabilidad de los autores.

\end{document}